\documentclass[10pt, journal]{IEEEtran}

\ifCLASSINFOpdf
\else
\fi
\usepackage{graphicx}
\usepackage{latexsym,amsmath,amssymb,bm}
\usepackage{booktabs}
\usepackage[utf8]{inputenc}
\usepackage[english]{babel}
\usepackage{url}
\usepackage{xcolor}
\usepackage[numbers]{natbib}

\begin{document}
\title{DeepLOB: Deep Convolutional Neural Networks for Limit Order Books}
\author{Zihao Zhang,~Stefan Zohren,~and~Stephen Roberts
\thanks{The authors are with the Oxford-Man Institute of Quantitative Finance, Department of Engineering Science, University of Oxford (e-mail: zihao@robots.ox.ac.uk). }
\thanks{Github: https://github.com/zcakhaa}}

\markboth{Journal of \LaTeX\ Class Files,~Vol.~XX, No.~XX, XXX}%
{Shell \MakeLowercase{\textit{et al.}}: Bare Demo of IEEEtran.cls for IEEE Journals}

\maketitle

\begin{abstract}
We develop a large-scale deep learning model to predict price movements from limit order book (LOB) data of cash equities. The architecture utilises convolutional filters to capture the spatial structure of the limit order books as well as LSTM modules to capture longer time dependencies. The proposed network outperforms all existing state-of-the-art algorithms on the benchmark LOB dataset \cite{ntakaris2017benchmark}. In a more realistic setting, we test our model by using one year market quotes from the London Stock Exchange and the model delivers a remarkably stable out-of-sample prediction accuracy for a variety of instruments. Importantly, our model translates well to instruments which were not part of the training set, indicating the model's ability to extract universal features. In order to better understand these features and to go beyond a ``black box'' model, we perform a sensitivity analysis to understand the rationale behind the model predictions and reveal the components of LOBs that are most relevant. The ability to extract robust features which translate well to other instruments is an important property of our model which has many other applications.
\end{abstract}


%
\IEEEpeerreviewmaketitle

\section{Introduction}
\label{introduction}

\IEEEPARstart{I}{n} today's competitive financial world more than half of the markets use electronic Limit Order Books (LOBs) \cite{parlour2008limit} to record trades \cite{rosu2010liquidity}. Unlike traditional quote-driven marketplaces, where traders can only buy or sell an asset at one of the prices made publicly by market makers, traders now can directly view all resting limit orders\footnote{Limit orders are orders that do not match immediately upon submission and are also called passive orders. This is opposed to orders that match immediately, so-called aggressive orders, such as a market order. A LOB is simply a record of all resting/outstanding limit orders at a given point in time.} in the limit order book of an exchange. Because limit orders are arranged into different levels based on their submitted prices, the evolution in time of a LOB represents a multi-dimensional problem with elements representing the numerous prices and order volumes/sizes at multiple levels of the LOB on both the buy and sell sides. 

A LOB is a complex dynamic environment with high dimensionality, inducing modelling complications that make traditional methods difficult to cope with. Mathematical finance is often dominated by models of evolving price sequences. This leads to a range of Markov-like models with stochastic driving terms, such as the vector autoregressive model (VAR) \cite{zivot2006vector} or the autoregressive integrated moving average model (ARIMA) \cite{ariyo2014stock}. These models, to avoid excessive parameter spaces, often rely on handcrafted features of the data. However, given the billions of electronic market quotes that are generated everyday, it is natural to employ more modern data-driven machine learning techniques to extract such features.

In addition, limit order data, like any other financial time-series data is notoriously non-stationary and dominated by stochastics. In particular, orders at deeper levels of the LOB are often placed and cancelled in anticipation of future price moves and are thus even more prone to noise. Other problems, such as auction and dark pools \cite{carrie2006new}, also add additional difficulties, bringing ever more unobservability into the environment. The interested reader is referred to \cite{gould2013limit} in which a number of these issues are reviewed. 

In this paper we design a novel deep neural network architecture that incorporates both convolutional layers as well as Long Short-Term Memory (LSTM) units to predict future stock price movements in large-scale high-frequency LOB data. One advantage of our model over previous research \cite{chiang2016adaptive} is that it has the ability to adapt for many stocks by extracting representative features from highly noisy data.

In order to avoid the limitations of handcrafted features, we use a so-called Inception Module \cite{szegedy2015going} to wrap convolutional and pooling layers together. The Inception Module helps to infer local interactions over different time horizons. The resulting feature maps are then passed into LSTM units which can capture dynamic temporal behaviour. We test our model on a publicly available LOB dataset, known as FI-2010 \cite{ntakaris2017benchmark}, and our method remarkably outperforms all existing state-of-the-art algorithms. However, the FI-2010 dataset is only made up of 10 consecutive days of down-sampled pre-normalised data from a less liquid market. While it is a valuable benchmark set, it is arguable not sufficient to fully verify the robustness of an algorithm. To ensure the generalisation ability of our model, we further test it by using one year order book data for 5 stocks from the London Stock Exchange (LSE). To minimise the problem of overfitting to backtest data, we carefully optimise any hyper-parameter on a separate validation set before moving to the out-of-sample test set. Our model delivers robust out-of-sample prediction accuracy across stocks over a test period of three months. 

As well as presenting results on out-of-sample data (in a timing sense) from stocks used to form the training set, we also test our model on out-of-sample (in both timing and data stream sense) stocks that are not part of the training set. Interestingly, we still obtain good results over the whole testing period. We believe this observation shows not only that the proposed model is able to extract robust features from order books, but also indicates the existence of universal features in the order book that modulate stock demand and price. The ability to transfer the model to new instruments opens up a number of possibilities that we consider for future work. 

To show the practicability of our model we use it in a simple trading simulation. We focus on sufficiently liquid stocks so that slippage and market impact are small. Indeed, these stocks are generally harder to predict than less liquid ones. Since our trading simulation is mainly meant as a method of comparison between models we assume trading takes place at mid-price\footnote{The  average of the best buy and best sell prices in the market at the time.} and compare gross profits before fees. The former assumption is equivalent to assuming that one side of the trade may be entered into passively and the latter assumes that different models trade similar volumes and would thus be subject to similar fees. Our focus here is using a simulation as a measure of the relative value of the model predictions in a trading setting. Under these simplifications, our model delivers significantly positive returns with a relatively small risk.

Although our network achieves good performance, a complex ``black box'' system, such as a deep neural network, has limited use for financial applications without some understanding of the rationale behind the model predictions. Here we exploit the model-agnostic LIME method \cite{ribeiro2016should} to highlight highly relevant components in the order book to gain a better understanding between our predictions and model inputs. Reassuringly, these conform to sensible (though arguably unusual) patterns of activity in both price and volume within the order book.

\paragraph*{Outline} The remainder of the paper is as follows. Section \ref{relatedwork} introduces background and related work. Section \ref{datasec} describes limit order data and the various stages of data preparation. We present our network architecture in Section \ref{modelsec} and give justifications behind each component of the model. In Section \ref{resultsec} we compare our work with a large group of popular methods. Section \ref{conclusion} summarises our findings and considers extensions and future work.


\section{Background and Related Work} \label{relatedwork}
Research on the predictability of stock markets has a long history in the financial literature e.g., \cite{ang2006stock, bacchetta2009predictability}. Although opinions differ regarding the efficiency of markets, many widely accepted studies show that financial markets are to some extent predictable \cite{bollerslev2014stock, ferreira2011forecasting, mandelbrot2007misbehavior, mandelbrot2008fractals}. Two major classes of work which attempt to forecast financial time-series are, broadly speaking, statistical parametric models and data-driven machine learning approaches \cite{agrawal2013state}. Traditional statistical methods generally assume that the time-series under study are generated from a parametric process \cite{cavalcante2016computational}. There is, however,  agreement that stock returns behave in more complex ways, typically highly nonlinearly \cite{cao2005comparison, sirignano2018universal}. Machine learning techniques are able to capture such arbitrary nonlinear relationships with little, or no, prior knowledge regarding the input data \cite{atsalakis2009surveying}.

Recently, there has been a surge of interest to predict limit order book data by using machine learning algorithms \cite{ntakaris2017benchmark, tran2017tensor, tran2017multilinear, passalis2018temporal, tran2018temporal, tsantekidis2017forecasting, tsantekidis2018using, sirignano2018universal, tsantekidis2017using, dixon2017classification}. Among many machine learning techniques, pre-processing or feature extraction is often performed as financial time-series data is highly stochastic. Generic feature extraction approches have been implemented, such as the Principal Component Analysis (PCA) and the Linear Discriminant Analysis (LDA) in the work of \cite{passalis2018temporal}. However these extraction methods are static pre-processing steps, which are not optimised to maximise the overall objective of the model that observes them. In the work of \cite{tran2018temporal, passalis2018temporal}, the Bag-of-Features model (BoF) is expressed as a neural layer and the model is trained end-to-end using the back-propagation algorithm, leading to notably better results on the FI-2010 dataset \cite{ntakaris2017benchmark}. These works suggest the importance of a data driven approach to extract representative features from a large amout of data. In our work, we advocate the end-to-end training and show that the deep neural network by itself not only leads to even better results but also transfers well to new instruments (not part of the training set) - indicating the ability of networks to extract ``universal'' features from the raw data.

Arguably, one of the key contributions of modern deep learning is the addition of feature extraction and representation as part of the learned model. The Convolutional Neural Network (CNN) \cite{lecun1995convolutional} is a prime example, in which information extraction, in the form of filter banks, is automatically tuned to the utility function that the entire network aims to optimise. CNNs have been successfully applied to various application domains, for example, object tracking \cite{wang2013learning}, object-detection \cite{girshick2014rich} and segmentation \cite{long2015fully}. However, there have been but a few published works that adopt CNNs to analyse financial microstructure data \cite{chen2016financial, doering2017convolutional, tsantekidis2017forecasting} and the existing CNN architectures are rather unsophisticated and lack of thorough investigation. Just like when moving from ``AlexNet'' \cite{krizhevsky2012imagenet} to ``VGGNet'' \cite{simonyan2014very}, we show that a careful design of network archiecture can lead to better results compared with all existing methods. 

The Long Short-Term Memory (LSTM) \cite{hochreiter1997long} was originally proposed to solve the vanishing gradients problem \cite{bengio1994learning} of recurrent neural networks, and has been largely used in applications such as language modelling \cite{sundermeyer2012lstm} and sequence to sequence learning \cite{sutskever2014sequence}. Unlike CNNs which are less widely applied in financial markets, the LSTM has been popular in recent years, \cite{bao2017deep, tsantekidis2017using, selvin2017stock, fischer2017deep, di2016artificial, dixon2017sequence, nelson2017stock, sirignano2018universal} all utilising LSTMs to analyse financial data. In particular, \cite{sirignano2018universal} uses limit order data from 1000 stocks to test a four layer LSTM model. Their results show a stable out-of-sample prediction accuracy across time, indicating the potential benefits of deep learning methods. To the best of our knowledge, there is no work that combines CNNs with LSTMs to predict stock price movements and this is the first extensive study to apply a nested CNN-LSTM model to raw market data. In particular, the usage of the Inception Model in this context is novel and is essential in inferring the optimal ``decay rates'' of the extracted features.


\section{Data, Normalisation and Labelling} \label{datasec}

\subsection{Limit Order Books}

We first introduce some basic definitions of limit order books (LOBs). For classical references on market microstructure the reader is referred to \cite{harris2003trading, o1995market} and for a short review on LOBs in particular we refer to \cite{gould2013limit}. Here we follow the conventions of \cite{gould2013limit}. A LOB has two types of orders: bid orders and ask orders. A bid (ask) order is an order to buy (sell) an asset at or below (above) a specified price. The bid orders have prices $\textbf{P}_b(t)$ and sizes/volumes $\textbf{V}_b(t)$, and the ask orders have prices $\textbf{P}_a(t)$ and sizes/volumes $\textbf{V}_a(t)$. Both $\textbf{P}(t)$ and $\textbf{V}(t)$ are vectors representing values at different price levels of an asset. 

\begin{figure}[!t]
\centering
\includegraphics[width=3in, height=3.1in]{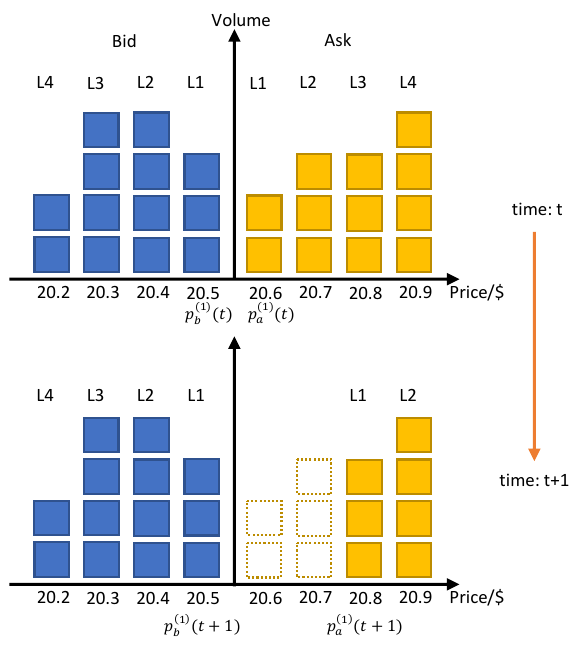}
\caption{A slice of LOB at time $t$ and $t+1$. L1 represents the respective first level, L2 the second, etc. $p_a^{(1)}(t)$ is the lowest ask price (best ask) and $p_b^{(1)}(t)$ is the highest bid price (best bid) at time $t$.}
\label{fig:lob}
\end{figure}

Figure~\ref{fig:lob} illustrates the above concepts. The upper plot shows a slice of a LOB at time $t$. Each square in the plot represents an order of nominal size 1. This is done for simplicity, in reality different orders can be of different sizes. The blue bars represent bid orders and the yellow bars represent ask orders. Orders are sorted into different levels based on their submitted prices, where L1 represents the first level and so on. Each level contains two values: price and volume. On the bid side, $\textbf{P}_b(t)$ and $\textbf{V}_b(t)$ are 4-vectors in this example. We use  $p_b^{(1)}(t)$ to denote the highest available price for a buying order (first bid level). Similarly, $p_a^{(1)}(t)$ is the lowest available selling order (first ask level). The bottom plot shows the action of an incoming market order to buy 5 shares at time $t+1$. As a result, the entire first and second ask-levels are executed against that order and $p_a^{(1)}(t+1)$ moved to 20.8 from 20.6 at time $t$.

\subsection{Input Data}

We test our model on two datasets: the FI-2010 dataset \cite{ntakaris2017benchmark} and one year length of limit order book data from the London Stock Exchange (LSE). The FI-2010 dataset \cite{ntakaris2017benchmark} is the first publicly available benchmark dataset of high-frequency limit order data and extracted time series data for five stocks from the Nasdaq Nordic stock market for a time period of 10 consecutive days. Many earlier algorithms are tested on this dataset and we use it to establish a fair comparison to other algorithms. However, 10 days is an insufficient amount of data to fully test the robustness and generalisation ability of an algorithm as the problem of overfitting to backtest data is severe and we often expect a signal to be consistent over a few months. 

To address the above concerns, we train and test our model on limit order book data of one year length for Lloyds Bank, Barclays, Tesco, BT and Vodafone. These five instruments are among the most liquid stocks listed on the London Stock Exchange. It is generally more difficult to train models on more liquid stocks, but at the same time, those instruments are easier to trade without price impact so making the simple trading simulation used to assess performance more realistic. The data includes all LOB updates for the above names. It spans all trading days from 3rd January 2017 to 24th December 2017 and we restrict it to the interval between 08:30:00 and 16:00:00, so that only normal trading activities occur and no auction takes place. Each state of the LOB contains 10 levels on each side and each level contains information on both price and volume. Therefore, we have a total of 40 features at each timestamp. Note that the FI-2010 dataset is actually downsampled limit order book data because the authors followed \cite{kercheval2015modelling} to create additional features by using every non-overlapping block of 10 events. We did not perform any processing on our data and only feed raw order book information to our algorithm.

Overall, our LSE dataset is made up of 12 months, and has more than 134 million samples. On average, there are 150,000 events per day per stock. The events are irregularly spaced in time. The time interval, $\Delta_{k, k+1}$, between two events can vary considerably from a fraction of a second to seconds, and $\Delta_{k, k+1}$ is on average 0.192 seconds in the dataset. We take the first 6 months as training data, the next 3 months as validation data and the last 3 months as test data. In the context of high-frequency data, 3 months test data corresponds to millions of observations and therefore provides sufficient scope for testing model performance and estimating model accuracy. 

\subsection{Data Normalisation and Labelling}

The FI-2010 dataset \cite{ntakaris2017benchmark} provides 3 different normalised dataset: $z$-score, min-max and decimal precision normalisation. We used data normalised by $z$-score without any emendation and found subtle difference when using the other two normalisation schemes. For the LSE dataset, we again use  standardisation ($z$-score) to normalise our data, but use the mean and standard deviation of the previous 5 days' data to normalise the current day's data (with a separate normalisation for each instrument).  We want to emphasize the importance of normalisation because the performance of machine learning algorithms often depends it. As financial time-series usually experiences regime shifts, using a static normalisation scheme is not appropriate for a dataset of one year length. The above  method is dynamic and the normalised data often falls into a reasonable range. We use the 100 most recent states of the LOB as an input to our model for both datasets. Specifically, a single input is defined as $X = [x_1, x_2, \cdots, x_t, \cdots, x_{100}]^T \in \mathbb{R}^{100 \times 40}$, where $x_t = [ p_a^{(i)}(t), v_a^{(i)}(t), p_b^{(i)}(t), v_b^{(i)}(t)]_{i=1}^{n=10}$. $p^{(i)}$ and $v^{(i)}$ denote the price and volume size at $i$-th level of a limit order book. 

After normalising the limit order data, we use the mid-price 
\begin{equation} \label{eq:2}
p_t = \frac{p_{a}^{(1)}(t) + p_{b}^{(1)}(t)}{2},
\end{equation}
to create labels that represent the direction of price changes. Although no order can transact exactly at the mid-price, it expresses a general market value for an asset and it is frequently quoted when we want a single number to represent an asset price.

Because financial data is highly stochastic, if we simply compare $p_t$ and $p_{t+k}$ to decide the price movement, the resulting label set will be noisy. In the works of \cite{ntakaris2017benchmark} and \cite{tsantekidis2017forecasting}, two smoothing labelling methods are introduced. We briefly recall the two methods here. First, let $m_-$ denote the mean of the previous $k$ mid-prices and $m_+$ denote the mean of the next $k$ mid-prices:
\begin{equation} \label{eq:3}
\begin{split}
m_-(t) = \frac{1}{k} \sum_{i=0}^k p_{t-i}\\
m_+(t) = \frac{1}{k} \sum_{i=1}^k p_{t+i}
\end{split}
\end{equation}
where $p_t$ is the mid-price defined in Equation \eqref{eq:2} and $k$ is the prediction horizon. Both methods use the percentage change ($l_t$) of the mid-price to decide directions. We can now define
\begin{equation} \label{eq:4}
l_t = \frac{m_+(t) - p_t}{p_t}
\end{equation}
\begin{equation} \label{eq:5}
l_t = \frac{m_+(t) - m_-(t)}{m_-(t)}
\end{equation}
Both are methods to define the direction of price movement at time $t$, where the former, Equation~\ref{eq:4}, was used in \cite{ntakaris2017benchmark} and the latter, Equation~\ref{eq:5}, in \cite{tsantekidis2017forecasting}.

\begin{figure}[!t]
\centering
\includegraphics[width=3.25in, height=1.5in]{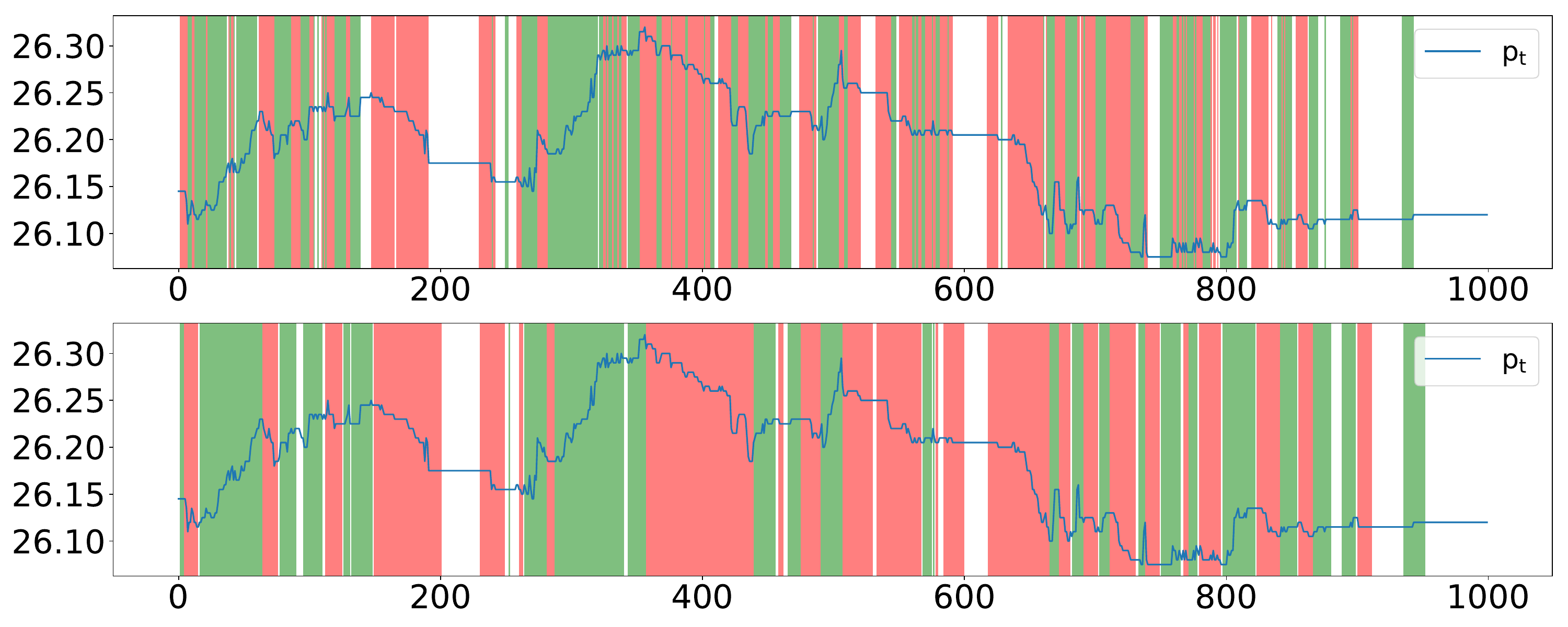}
\caption{An example of two smoothed labelling methods based on a same threshold ($\alpha$) and same prediction horizon ($k$). Green shading represents a +1 signal and red a -1. \textbf{Top:} \cite{ntakaris2017benchmark}'s method and \textbf{Bottom:} \cite{tsantekidis2017forecasting}'s method.}
\label{fig:smooth_label}
\end{figure}

The labels are then decided based on a threshold ($\alpha$) for the percentage change ($l_t$). If $l_t > \alpha$ or $l_t < -\alpha$, we define it as up ($+1$) or down ($-1$). For anything else, we consider it as stationary ($0$). Figure~\ref{fig:smooth_label} provides a graphical illustration of two labelling methods on the same threshold ($\alpha$) and the same prediction horizon ($k$). All the labels classified as down ($-1$) are shown as red areas and up ($+1$) as green areas. The uncoloured (white) regions correspond to stationary ($0$) labels. 

The FI-2010 dataset \cite{ntakaris2017benchmark} adopts the method in Equation~\ref{eq:4} and we directly used their labels for fair comparison to other methods. However, the produced labels are less consistent as shown on the top of Figure~\ref{fig:smooth_label} because this method fits closer to real prices as smoothing is only applied to future prices. This is essentially detrimental for designing trading algorithms as signals are not consistent here leading to many redundant trading actions thus incurring larger transaction costs. 

Further, the FI-2010 dataset was collected in 2010 and the instruments were less liquid compared to now. We experimented with this approach in \cite{ntakaris2017benchmark} on our data from the London Stock Exchange and found the resulting labels are rather stochastic, therefore we adopt the method in Equation~\ref{eq:5} for our LSE dataset to produce more consistent signals.

\section{Model Architecture} \label{modelsec}

\subsection{Overview}
\begin{figure}[!t]
\centering
\includegraphics[width=1.2in, height=3.1in]{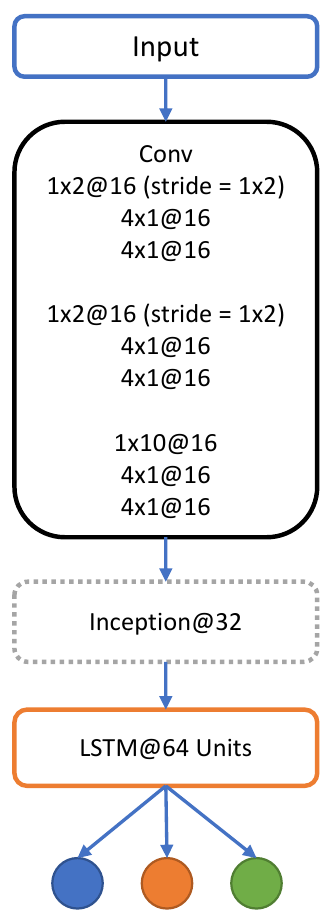}
\caption{Model architecture schematic. Here 1x2@16 represents a convolutional layer with 16 filters of size ($1 \times 2$). `1' convolves through time indices and `2' convolves different limit order book levels.}
\label{fig:simple_model}
\end{figure}

We here detail our network architecture, which comprises three main building blocks: standard convolutional layers, an Inception Module and a LSTM layer, as shown in Figure~\ref{fig:simple_model}. The main idea of using CNNs and Inception Modules is to automate the process of feature extraction as it is often difficult in financial applications since financial data is notoriously noisy with a low signal-to-noise ratio. Technical indicators such as MACD and the Relative Strength Index are included as inputs and preprocessing mechanisms such as principal component analysis (PCA) \cite{abraham2001hybrid} are often used to transform raw inputs. However, none of these processes is trivial, they make tacit assumptions and further, it is questionable if financial data can be well-described with parametric models with fixed parameters. In our work, we only require the history of LOB prices and sizes as inputs to our algorithm. Weights are learned during inference and features, learned from a large training set, are data-adaptive, removing the above constraints. A LSTM layer is then used to capture additional time dependencies among the resulting features. We note that very short time-dependencies are already captured in the convolutional layer which takes ``space-time images'' of the LOB as inputs.

\subsection{Details of Each Component}
\paragraph{Convolutional Layer}

Recent development of electronic trading algorithms often submit and cancel vast numbers of limit orders over short periods of time as part of their trading strategies \cite{hendershott2011does}. These actions often take place deep in a LOB and it is seen \cite{gould2013limit} that more than 90\% of orders end in cancellation rather than matching, therefore practitioners consider levels further away from best bid and ask levels to be less useful in any LOB. In addition, the work of \cite{cao2009information} suggests that the best ask and best bid (L1-Ask and L1-Bid) contribute most to the price discovery and the contribution of all other levels is considerably less, estimated at as little as 20\%. As a result, it would be otiose to feed all level information to a neural network as levels deep in a LOB are less useful and can potentially even be misleading.
Naturally, we can smooth these signals by summarising the information contained in deeper levels. We note that convolution filters used in any CNN architecture are discrete convolutions, or finite impulse response (FIR) filters, from the viewpoint of signal processing \cite{orfanidis1995introduction}. FIR filters are popular smoothing techniques for denoising target signals and they are simple to implement and work with. We can write any FIR filter in the following form:
\begin{equation}
y(n) = \sum_{k=0}^M b_k x(n-k)
\end{equation}
where the output signal $y(n)$ at any time is a weighted sum of a finite number of past values of the input signal $x(n)$. The filter order is denoted as $M$ and $b_k$ is the filter coefficient. In a convolutional neural network, the coefficients of the filter kernel are not obtained via a statistical objective from traditional signal filtration theory, but are left as degrees of freedom which the network infers so as to extremise its value function at output. 

The details of the first convolutional layer inevitably need some consideration. As convolutional layers  operate a small kernel to ``scan'' through input data, the layout of limit order book information is vital. Recall that we take the most 100 recent updates of an order book to form a single input and there are 40 features per time stamp, so the size of a single input is $(100  \times  40)$. We organise the 40 features as following:
\begin{equation}
\{  p_a^{(i)}(t), v_a^{(i)}(t), p_b^{(i)}(t), v_b^{(i)}(t) \}_{i=1}^{n=10}
\end{equation}
where $i$ denotes the $i$-th level of a limit order book. The size of our first convolutional filter is $(1 \times 2)$ with stride of $(1 \times 2)$. The first layer essentially summarises information between price and volume $\{p^{(i)}, v^{(i)}\}$ at each order book level. The usage of stride is necessary here as an important property of convolutional layers is parameter sharing. This property is attractive as less parameters are estimated, largely avoiding overfitting problems. However, without strides, we would apply same parameters to $\{p^{(i)}, v^{(i)}\}$ and $\{v^{(i)}, p^{(i+1)}\}$. In other words, $p^{(i)}$ and $v^{(i)}$ would share same parameters because the kernel filter moves by one step, which is obviously wrong as price and volume form different dynamic behaviors.

Because the first layer only captures information at each order book level, we would expect representative features to be extracted when integrating information across multiple order book levels. We can do this by utilising another convolutional layer with filter size $(1 \times 2)$ and stride $(1 \times 2)$. The resulting feature maps actually form the micro-price defined by \cite{gatheral2010zero}:
\begin{equation}
\begin{split}
    p^{\mathrm{micro \ price}} &= Ip_a^{(1)} + (1-I)p_b^{(1)} \\
    I &= \frac{v_b^{(1)}}{v_a^{(1)} + v_b^{(1)}}
\end{split}
\end{equation}
The weight $I$ is called the imbalance. The micro-price is an important indicator as it considers volumes on bid and ask side, and the imbalance between bid and ask size is a very strong indicator of the next price move. This feature of imbalances has been reported by a variety of researchers \cite{nevmyvaka2006reinforcement, avellaneda2011forecasting, burlakov2012optimal, harris2013maker, lipton2013trade}. Unlike the micro-price where only the first order book level is considered, we utilise convolutions to form micro-prices for all levels of a LOB so the resulting features maps are of size $(100, 10)$ after two layers with strides. Finally, we integrate all information by using a large filter of size $(1 \times 10)$ and the dimension of our feature maps before the Inception Module is $(100, 1)$. 

We apply zero padding to every convolutional layer so the time dimension of our inputs does not change and Leaky Rectifying Linear Units (Leaky-ReLU) \cite{maas2013rectifier} are used as activation functions. The hyper-parameter (the small gradient when the unit is not active) of the Leaky-ReLU is set to 0.01, evaluated by grid search on the validation set. 

Another important property of convolution is that of equivariance to translation \cite{Goodfellow-et-al-2016}. Specifically, a function $f(x)$ is equivariant to a function $g$ if $f(g(x)) = g(f(x))$. For example, suppose that there exists a main classification feature $m$ located at $(x_m, y_m)$ of an image $I(x,y)$. If we shift every pixel of $I$ one unit to the right, we get a new image $I'$ where $I'(x,y) = I(x-1,y)$. We can still obtain the main classification feature $m'$ in $I'$ and $m = m'$, while the location of $m'$ will be at $(x_{m'}, y_{m'}) = (x_m-1, y_m)$. This is important to time-series data, because convolution can find universal features that are decisive to final outputs. In our case, suppose a feature that studies imbalance is obtained at time $t$. If the same event happens later at time $t'$ in the input, the exact feature can be extracted later at $t'$.

We do not use any pooling layer except in the Inception Modules. Although pooling layers help us find representations invariant to translations of the input, the smoothing nature of pooling can cause under-fitting. Common pooling layers are designed for image processing tasks, and they are most powerful when we only care if certain features exist in the inputs instead of where they exist \cite{Goodfellow-et-al-2016}. Time-series data has different characteristics from images and the location of representative features is important. Our experiences show that pooling layers in the convolutional layer, at least, cause under-fitting problems to the LOB data. However, we think pooling is important and new pooling methods should be designed to process time-series data as it is a promising solution to extract invariant features. \\

\paragraph{Inception Module}

We note that all filters of a standard convolutional layer have fixed size. If, for example, we employ filters of size ($4 \times 1$), we capture local interactions amongst data over four time steps. However, we can capture dynamic behaviours over multiple timescales by using Inception Modules to wrap several convolutions together. We find that this offers a performance improvement to the resultant model. 

The idea of the Inception Module can be also considered as using different moving averages in technical analysis. Practitioners often use moving averages with different decay weights to observe time-series momentum \cite{moskowitz2012time}. If a large decay weight is adopted, we get a smoother time-series that well represents the long-term trend, but we could miss small variations that are important in high-frequency data. In practice, it is a daunting task to set the right decay weights. Instead, we can use Inception Modules and the weights are then learned during back-propagation. 

\begin{figure}[!t]
\centering
\includegraphics[width=3in, height=2in]{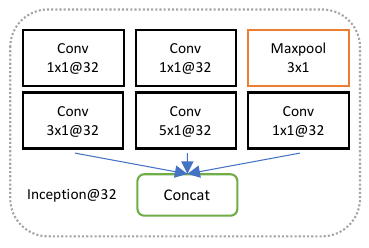}
\caption{The Inception Module used in the model. For example, $3 \times 1 @32$ represents a convolutional layer with 32 filters of size ($3 \times 1$).}
\label{fig:incep}
\end{figure}

In our case, we split the input into a small set of lower-dimensional representations by using $1 \times 1$ convolutions, transform the representations by a set of filters, here $3 \times 1$ and $5 \times 1$, and then merge the outputs. A max-pooling layer is used inside the Inception Module, with stride 1 and zero padding. ``Inception@32" represents one module and indicates all convolutional layers have 32 filters in this module, and the approach is depicted schematically in Figure \ref{fig:incep}. The $1 \times 1$ convolutions form the Network-in-Network approach proposed in \cite{lin2013network}. Instead of applying a simple convolution to our data, the Network-in-Network method uses a small neural network to capture non-linear properties of our data. We find this method to be effective and it gives us an improvement on prediction accuracy.

\paragraph{LSTM Module and Output}
In general, a fully connected layer is used to classify the input data. However, all inputs to the fully connected layer are assumed independent of each other unless multiple fully connected layers are used. Due to the usage of Inception Module in our work, we have a large number of features at end. Just using one fully connected layer with 64 units would result in more than 630,000 parameters to be estimated, not to mention multiple layers. In order to capture temporal relationship that exist in the extracted features, we replace the fully connected layers with LSTM units. The activation of a LSTM unit is fed back to itself and the memory of past activations is kept with a separate set of weights, so the temporal dynamics of our features can be modelled. We use 64 LSTM units in our work, resulting in about 60,000 parameters, leading to 10 times fewer parameters to be estimated. The last output layer uses a softmax activation function and hence the final output elements represent the probability of each price movement class at each time step. 


\section{Experimental Results} \label{resultsec}

\subsection{Experiments Settings}

We apply the same architecture to all our experiments in this section and the proposed model is denoted as DeepLOB. We learn the parameters by minimising the categorical cross-entropy loss. The Adaptive Moment Estimation algorithm, ADAM \cite{kingma2014adam}, is utilised and we set the parameter ``epsilon'' to 1 and the learning rate to 0.01. The learning is stopped when validation accuracy does not improve for 20 more epochs. This is about 100 epochs for the FI-2010 dataset and 40 epochs for the LSE dataset. 

We train with mini-batches of size 32. We choose a small mini-batch size due to the findings in \cite{keskar2016large} in which they suggest that large-batch methods tend to converge to narrow deep minima of the training functions, but small-batch methods consistently converge to shallow broad minima. All models are built using Keras \cite{chollet2015keras} based on the TensorFlow backend \cite{tensorflow2015-whitepaper}, and we train them using a single NVIDIA Tesla P100 GPU. 

\subsection{Experiments on the FI-2010 Dataset}

There are two experimental setups using the FI-2010 dataset. Following the convention of \cite{passalis2018temporal}, we denote them as Setup 1 and Setup 2.  Setup 1 splits the dataset into 9 folds based on a day basis (a standard anchored forward split). In the $i$-th fold, we train our model on the first $i$ days and test it on the $(i+1)$-th day where $i=1, \cdots, 9$. The second setting, Setup 2, originates from the works \cite{tsantekidis2017forecasting, tsantekidis2017using,tsantekidis2018using, tran2018temporal}  in which deep network architectures were evaluated. As deep learning techniques often require a large amount of data to calibrate weights, the first 7 days are used as the train data and the last 3 days are used as the test data in this setup. We evaluate our model in both setups here. 

Table~\ref{table:benchmark_transfer} shows the results of our model compared to other methods in Setup 1. Performance is measured by calculating the mean accuracy, recall, precision, and F1 score over all folds. As the FI-2010 dataset is not well balanced, \cite{ntakaris2017benchmark} suggests to focus on F1 score performance as fair comparisons. We have compared our model to all existing experimental results including Ridge Regression (RR) \cite{ntakaris2017benchmark}, Single-Layer-Feedforward Network (SLFN) \cite{ntakaris2017benchmark}, Linear Discriminant Analysis (LDA) \cite{tran2017tensor}, Multilinear Discriminant Analysis (MDA) \cite{tran2017tensor}, Multilinear Time-series Regression (MTR) \cite{tran2017tensor}, Weighted Multilinear Time-series Regression (WMTR) \cite{tran2017tensor}, Multilinear Class-specific Discriminant Analysis (MCSDA) \cite{tran2017multilinear}, Bag-of-Feature (BoF) \cite{passalis2018temporal}, Neural Bag-of-Feature (N-BoF) \cite{passalis2018temporal}, and Attention-augmented-Bilinear-Network with one hidden layer (B(TABL)) and two hidden layers (C(TABL)) \cite{tran2018temporal}. More methods such as PCA and Autoencoder (AE) are actually tested in their works but, for simplicity, we only report their best results and our model achieves better performance. 

However, the Setup 1 is not ideal for training deep learning models as we mentioned that deep network often requires a large amount of data to calibrate weights. This anchored forward setup leads to only one or two days' training data for the first few folds and we observe worse performance in the first few days. As training data grows, we observe remarkably better results as shown in Table~\ref{table:benchmark} which shows the results of our network compared to other methods in Setup 2. In particular, the important difference between our model and CNN-I \cite{tsantekidis2017forecasting} and CNN-II \cite{tsantekidis2018using} is due to network architecture and we can see huge improvements on performance here. In Table~\ref{table:computation_cost}, we compare the parameter sizes of DeepLOB with CNN-I \cite{tsantekidis2017forecasting}. Although our model has many more layers, there are far fewer parameters in our network due to the usage of LSTM layers instead of fully connected layers. 

We also report the computation time (forward pass) in milliseconds (ms) for available algorithms in Table~\ref{table:computation_cost}. Due to the development of GPUs, training deep networks is now feasible and it is swift to make predictions, making it possible for high frequency trading. We will discuss this more in the next section. 
\begin{table}[!t]
\caption{Setup 1: Experiment Results for the FI-2010 Dataset}
\label{table:benchmark_transfer}
\centering
\begin{tabular}{l|llll}
\toprule
Model    & Accuracy \% & Precision \%& Recall \% & F1 \%\\
\midrule
\multicolumn{5}{c}{Prediction Horizon k = 10} \\
\midrule
RR \cite{ntakaris2017benchmark}   &48.00     &41.80      &43.50  &41.00    \\
SLFN \cite{ntakaris2017benchmark} &64.30     &51.20      &36.60  &32.70    \\
LDA \cite{tran2017tensor}         &63.83     &37.93      &45.80  &36.28    \\
MDA \cite{tran2017tensor}         &71.92     &44.21      &60.07  &46.06    \\
MCSDA \cite{tran2017multilinear}  &83.66     &46.11      &48.00  &46.72    \\
MTR \cite{tran2017tensor}         &86.08     &51.68      &40.81  &40.14    \\
WMTR \cite{tran2017tensor}        &81.89     &46.25      &51.29  &47.87    \\
BoF \cite{passalis2018temporal}   &57.59     &39.26      &51.44  &36.28    \\
N-BoF \cite{passalis2018temporal} &62.70     &42.28      &61.41  &41.63    \\
B(TABL) \cite{tran2018temporal}   &73.62     &66.16      &68.81  &67.12    \\
C(TABL) \cite{tran2018temporal}   &78.01     &72.03      &74.04  &72.84    \\
DeepLOB                          &78.91     &78.47      &78.91  &\textbf{77.66}    \\
\midrule
\multicolumn{5}{c}{Prediction Horizon k = 50} \\
\midrule
RR \cite{ntakaris2017benchmark}   &43.90     &43.60      &43.30  &42.70    \\
SLFN \cite{ntakaris2017benchmark} &47.30     &46.80      &46.40  &45.90    \\
BoF \cite{passalis2018temporal}   &50.21     &42.56      &49.57  &39.56    \\
N-BoF \cite{passalis2018temporal} &56.52     &47.20      &58.17  &46.15    \\
B(TABL) \cite{tran2018temporal}   &69.54     &69.12      &68.84  &68.84   \\
C(TABL) \cite{tran2018temporal}   &74.81     &74.58      &74.27  &74.32    \\
DeepLOB                          &75.01     &75.10      &75.01  &\textbf{74.96}    \\
\midrule
\multicolumn{5}{c}{Prediction Horizon k = 100} \\
\midrule
RR \cite{ntakaris2017benchmark}   &42.90     &42.90      &42.90  &41.60    \\
SLFN \cite{ntakaris2017benchmark} &47.70     &45.30      &43.20  &41.00    \\
BoF \cite{passalis2018temporal}   &50.97     &42.48      &47.84  &40.84    \\
N-BoF \cite{passalis2018temporal} &56.43     &47.27      &54.99  &46.86    \\
B(TABL) \cite{tran2018temporal}   &69.31     &68.95      &69.41  &68.86    \\
C(TABL) \cite{tran2018temporal}   &74.07     &73.51      &73.80  &73.52    \\
DeepLOB                          &76.66     &76.77      &76.66  &\textbf{76.58}    \\
\bottomrule
\end{tabular}
\end{table}

\begin{table}[!t]
\caption{Setup 2: Experiment Results for the FI-2010 Dataset}
\label{table:benchmark}
\centering
\begin{tabular}{l|llll}
\toprule
Model    & Accuracy \% & Precision \% & Recall \%& F1 \% \\
\midrule
\multicolumn{5}{c}{Prediction Horizon k = 10} \\
\midrule
SVM \cite{tsantekidis2017using}      &-          &39.62      &44.92      &35.88    \\
MLP \cite{tsantekidis2017using}      &-          &47.81      &60.78      &48.27    \\
CNN-I \cite{tsantekidis2017forecasting}&-          &50.98      &65.54      &55.21    \\
LSTM \cite{tsantekidis2017using}     &-          &60.77      &75.92      &66.33    \\
CNN-II \cite{tsantekidis2018using}   &-          &56.00      &45.00      &44.00    \\
B(TABL) \cite{tran2018temporal}      &78.91      &68.04      &71.21      &69.20    \\
C(TABL) \cite{tran2018temporal}      &84.70      &76.95      &78.44      &77.63    \\
DeepLOB                             &84.47      &84.00      &84.47      &\textbf{83.40} \\
\midrule
\multicolumn{5}{c}{Prediction Horizon k = 20} \\
\midrule
SVM \cite{tsantekidis2017using}      &-          &45.08      &47.77      &43.20    \\
MLP \cite{tsantekidis2017using}      &-          &51.33      &65.20      &51.12    \\
CNN-I \cite{tsantekidis2017forecasting}&-         &54.79      &67.38      &59.17    \\
LSTM \cite{tsantekidis2017using}     &-          &59.60      &70.52      &62.37    \\
CNN-II \cite{tsantekidis2018using}   &-          &-      &-      &-    \\
B(TABL) \cite{tran2018temporal}      &70.80      &63.14      &62.25      &62.22    \\
C(TABL) \cite{tran2018temporal}      &73.74      &67.18      &66.94      &66.93    \\
DeepLOB                             &74.85      &74.06      &74.85      &\textbf{72.82}    \\
\midrule
\multicolumn{5}{c}{Prediction Horizon k = 50} \\
\midrule
SVM \cite{tsantekidis2017using}      &-          &46.05      &60.30      &49.42    \\
MLP \cite{tsantekidis2017using}      &-          &55.21      &67.14      &55.95    \\
CNN-I \cite{tsantekidis2017forecasting}&-          &55.58      &67.12      &59.44    \\
LSTM \cite{tsantekidis2017using}     &-          &60.03      &68.58      &61.43    \\
CNN-II \cite{tsantekidis2018using}   &-          &56.00      &47.00      &47.00    \\
B(TABL) \cite{tran2018temporal}      &75.58      &74.58      &73.09      &73.64    \\
C(TABL) \cite{tran2018temporal}      &79.87      &79.05      &77.04      &78.44    \\
DeepLOB                             &80.51      &80.38      &80.51      &\textbf{80.35}   \\
\bottomrule
\end{tabular}
\end{table}

\begin{table}[!t]
\caption{Average Computation Time of State-Of-The-Art Models}
\label{table:computation_cost}
\centering
\begin{tabular}{lll}
\toprule
Models  & Forward (ms) & Number of parameters \\
\midrule
BoF \cite{passalis2018temporal}      			& 0.972        & 86k                  \\
N-BoF \cite{passalis2018temporal}    		& 0.524        & 12k                  \\
CNN-I \cite{tsantekidis2017forecasting}	& 0.025        & 768k                 \\
LSTM \cite{tsantekidis2017using}     			& 0.061        & -                    \\
C(TABL) \cite{tran2018temporal} 							& 0.229        & -                     \\
DeepLOB                            								& 0.253        & 60k                    \\
\bottomrule
\end{tabular}
\end{table}

\subsection{Experiments on the London Stock Exchange (LSE)}

As we suggested, the FI-2010 dataset is not sufficient to verify a prediction model - it is far too short, downsampled and taken from a less liquid market. To perform a meaningful evaluation that can hold up to modern applications, we further test our method on stocks from the LSE of one year length with a testing period of three months. As mentioned in Section \ref{datasec}, we train our model on five stocks: Lloyds Bank (LLOY), Barclays (BARC), Tesco (TSCO), BT and Vodafone (VOD).  Recent work of \cite{sirignano2018universal} suggests that deep learning techniques can extract universal features for limit order data. To test this universality, we directly apply our model to five more stocks that were not part of the training data set \textbf{(transfer learning)}. We select HSBC, Glencore (GLEN), Centrica (CNA), BP and ITV for transfer learning because they are also among the most liquid stocks in the LSE. The testing period is the same three months as before, and the classes are roughly balanced.

\begin{figure}[!t]
\centering
\includegraphics[width=3.5in, height=1.1in]{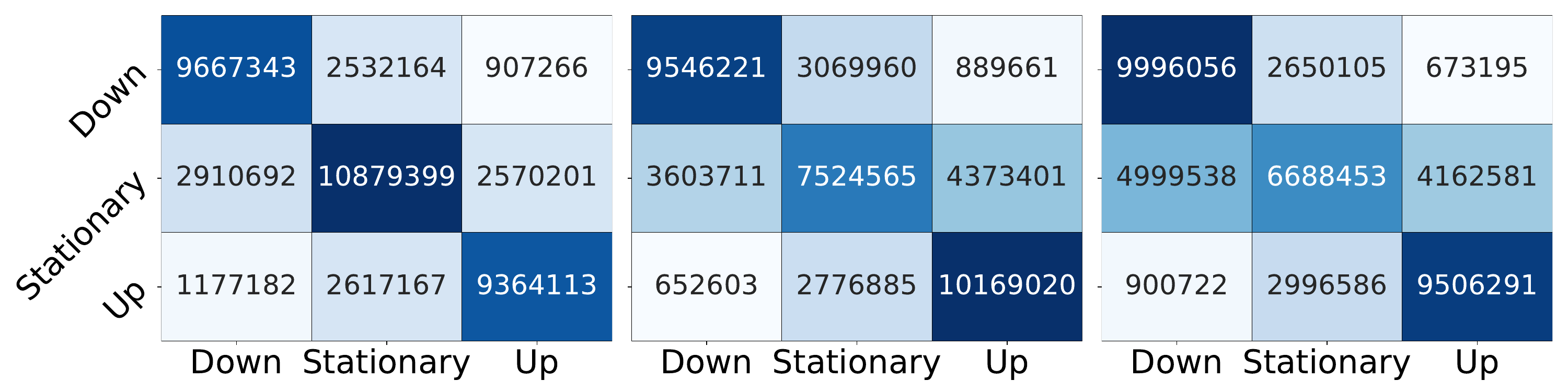}
\includegraphics[width=3.5in, height=1.1in]{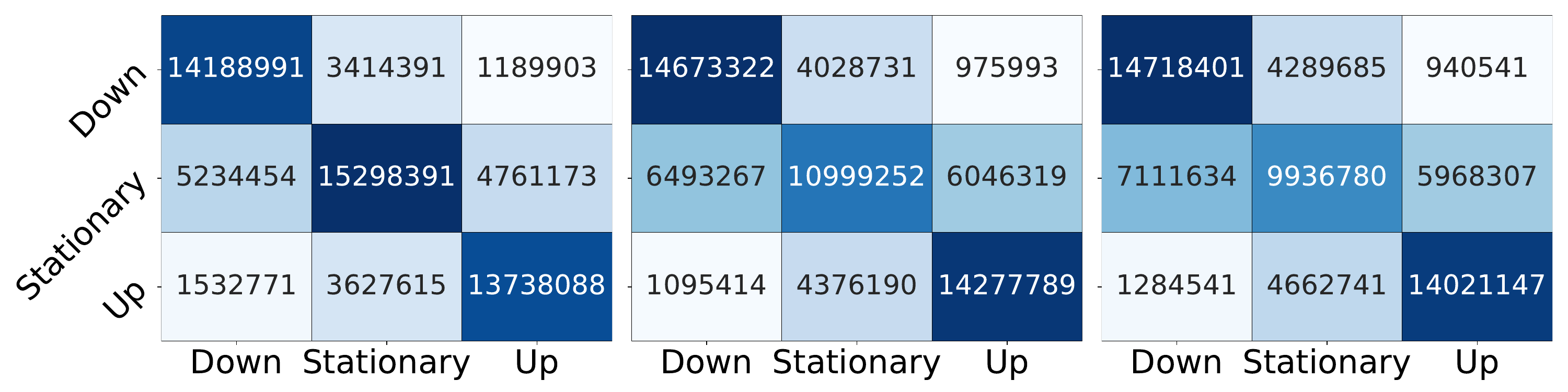}
\caption{Confusion matrices. \textbf{Top:} results on LLOY, BARC, TSCO, BT and VOD. From the left to right, prediction horizon $(k)$ equals 20, 50 and 100; \textbf{Bottom:} results on transfer learning (GLEN, HSBC, CNA, BP, ITV).}
\label{fig:confusion_matrix}
\end{figure}

\begin{figure}[!t]
\centering
\includegraphics[width=3.55in, height=1.5in]{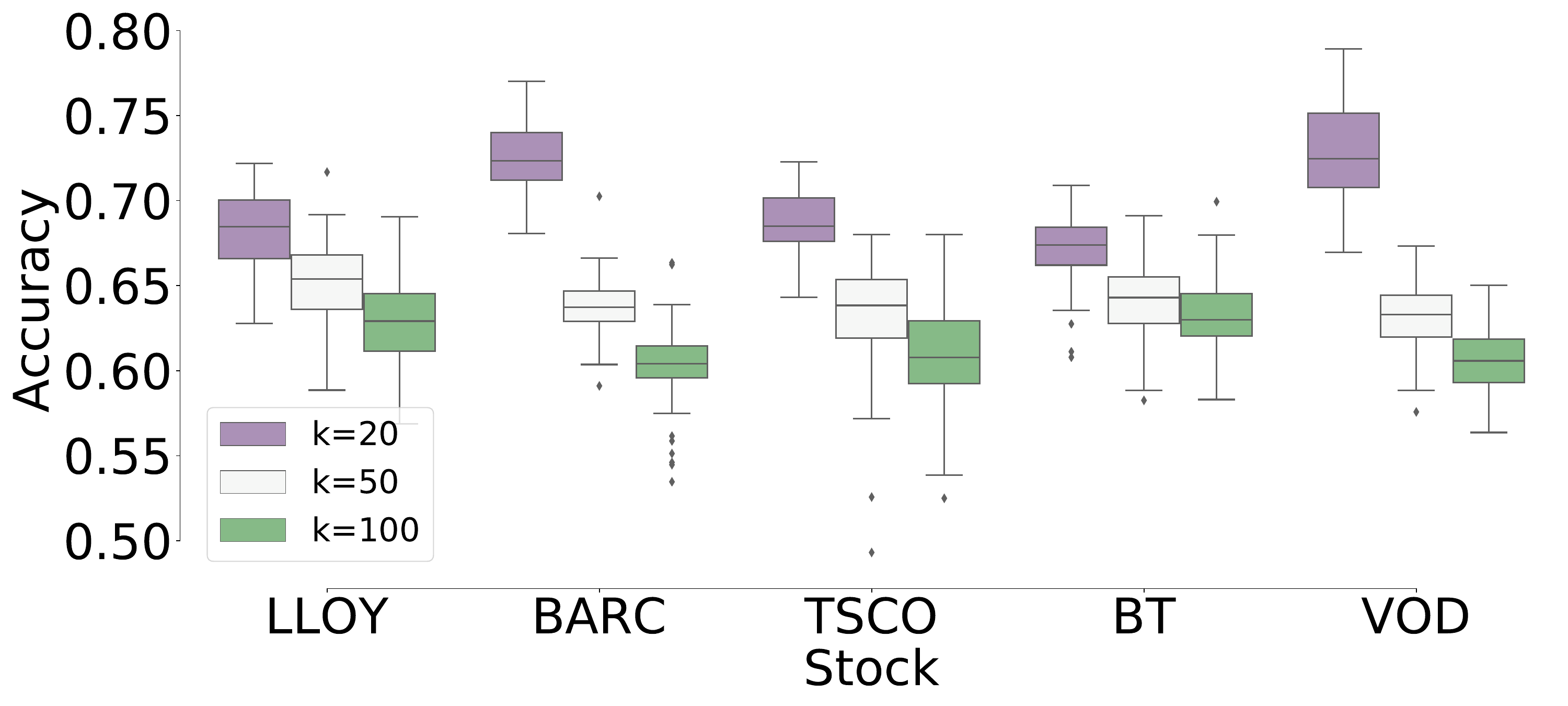}
\includegraphics[width=3.55in, height=1.4in]{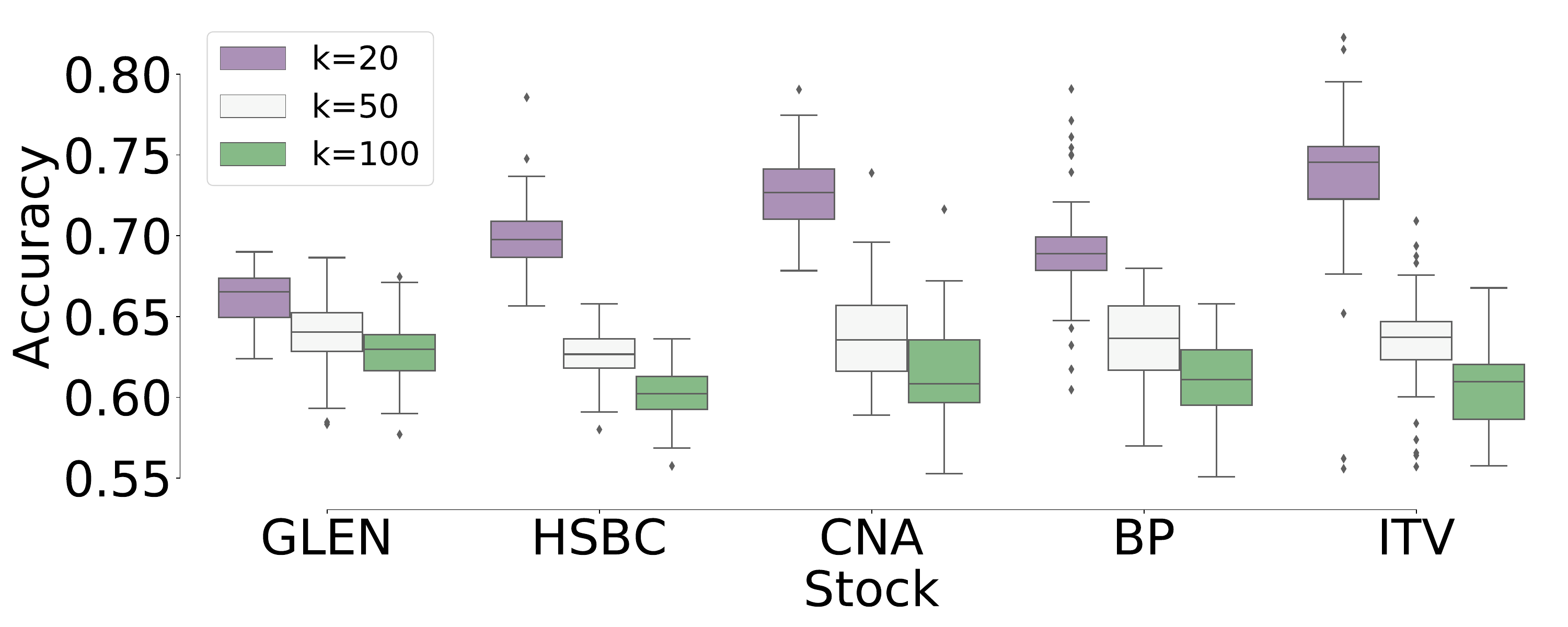}
\caption{Boxplots of daily accuracy for the different prediction horizons. \textbf{Top:} results on LLOY, BARC, TSCO, BT and VOD; \textbf{Bottom:} results on transfer learning (GLEN, HSBC, CNA, BP, ITV).}
\label{fig:boxplot_acc}
\end{figure}

Table~\ref{table:lse_result} presents the results of our model for all stocks on different prediction horizons. To better investigate the results, we display the confusion matrices in Figure~\ref{fig:confusion_matrix} and calculate the accuracy for every day and for every stock across the testing period. We use the boxplots in Figure~\ref{fig:boxplot_acc} to present this information and we can observe consistent and robust performance, with narrow interquartile range (IQR) and few outliers, for all stocks across the testing period. The ability of our model that generalises well to data not in the training set indicates that the CNN block in the algorithms, acting to extract features from the LOB, can capture universal patterns that relate to the price formation mechanism. We find this observation most interesting.

\begin{table}[!t]
\caption{Experiment Results for the LSE Dataset}
\label{table:lse_result}
\centering
\begin{tabular}{l|llll}
\toprule
Prediction Horizon & Accuracy \% & Precision \% & Recall  \% & F1 \% \\
\midrule
\multicolumn{5}{c}{Results on LLOY, BARC, TSCO, BT and VOD} \\ 
\midrule
k=20               &70.17          &70.17           &70.17        &70.15    \\
k=50              &63.93          &63.43           &63.93        &63.49    \\
k=100              &61.52          &60.73          &61.52       &60.65    \\
\midrule
\multicolumn{5}{c}{Results on Transfer Learning (GLEN, HSBC, CNA, BP, ITV)} \\ 
\midrule
k=20               &68.62          &68.64           &68.63        &68.48    \\
k=50              &63.44          &62.81           &63.45        &62.84    \\
k=100              &61.46         &60.68           &61.46        &60.77    \\
\bottomrule
\end{tabular}
\end{table}

\subsection{Performance of the Model in a Simple Trading Simulation}

A simple trading simulation is designed to test the practicability of our results. We set the number of shares per trade, $\mu$, to one both for simplicity and to minimise the market impact, ensuring orders to be executed at the best price. Although $\mu$ can be optimised to maximise the returns, for example, prediction probabilities are used to size the orders in \cite{zhang2018bdlob}, we would like to show that our algorithm can work even under this simple set-up.

To reduce the number of trades, we use following rules to take actions. At each time-step, our model generates a signal from the network outputs $(-1, 0,+1)$ to indicate the price movements in $k$ steps. Signals $(-1,0,+1)$ correspond to actions (sell, wait and buy). Suppose our model produces a prediction of $+1$ at time $t$, we then buy $\mu$ shares at time $t+5$ (taking slippage into account), and hold until $-1$ appears to sell all $\mu$ shares (we do nothing if $0$ appears). We apply the same rule to short selling and repeat the process during a day. All positions are closed by the end of the day, so we hold no stocks overnight. We make sure no trades take place at the time of auction, so no abnormal profits are generated. 

As the focus of our work is on predictions and the above simple simulation is a way of showing that this prediction is in principle monetisable. In particular, our aim is not to present a fully developed, stand-alone trading strategy. Realistic high-frequency strategies often require a combination of various trading signals in particular to time the exact entry and exit points of the trade. For the purpose of the above simulation we use mid-prices without transaction costs. While in particular the second assumption is not a reasonable assumption for a standalone strategy, we argue that (i) it is enough for a relative comparison of the above models and (ii) it is a good indicator of the relative value of the above predictor to a more complex high-frequency trading model. Regarding the first assumption, a mid-mid simulation, we note that in high-frequency trading, many participants are involved in market making, as it is difficult to design profitable fully aggressive strategies with such short holding periods. If we assume that we are able to enter the trade passively, while we exit it aggressively, crossing the spread, then this is effectively equivalent to a mid-mid trade. Such a situation arises naturally for example in investment banks which are involved in client market making. Regarding the second assumption, careful timing of the entry points as well as more elaborate trading rules, such as including position upsizing, should be able to account for additional profits to cover the transaction costs. In any case, as merely a metric of testing predictability of our model, the above simple simulation suffices.

\begin{figure*}[t]
\centering
\includegraphics[height=2.5in]{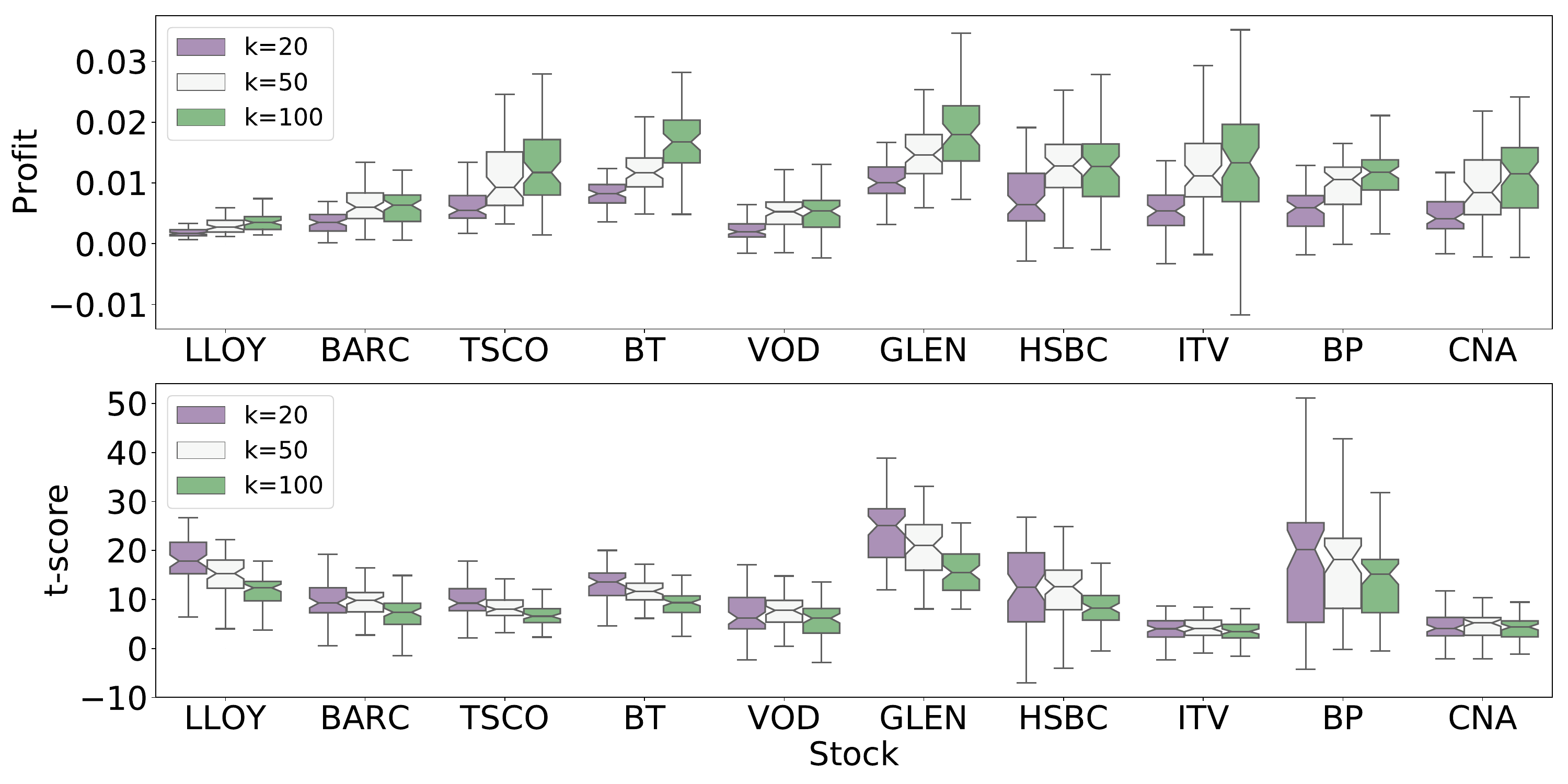}
\caption{Boxplots for normalised daily profits and $t$-statistics for different stocks and prediction horizons $(k)$. Profits are in GBX ($=\text{GBP}/100$).}
\label{fig:boxplot_profit}
\end{figure*}

\begin{figure*}[t]
\centering
\includegraphics[height=2in]{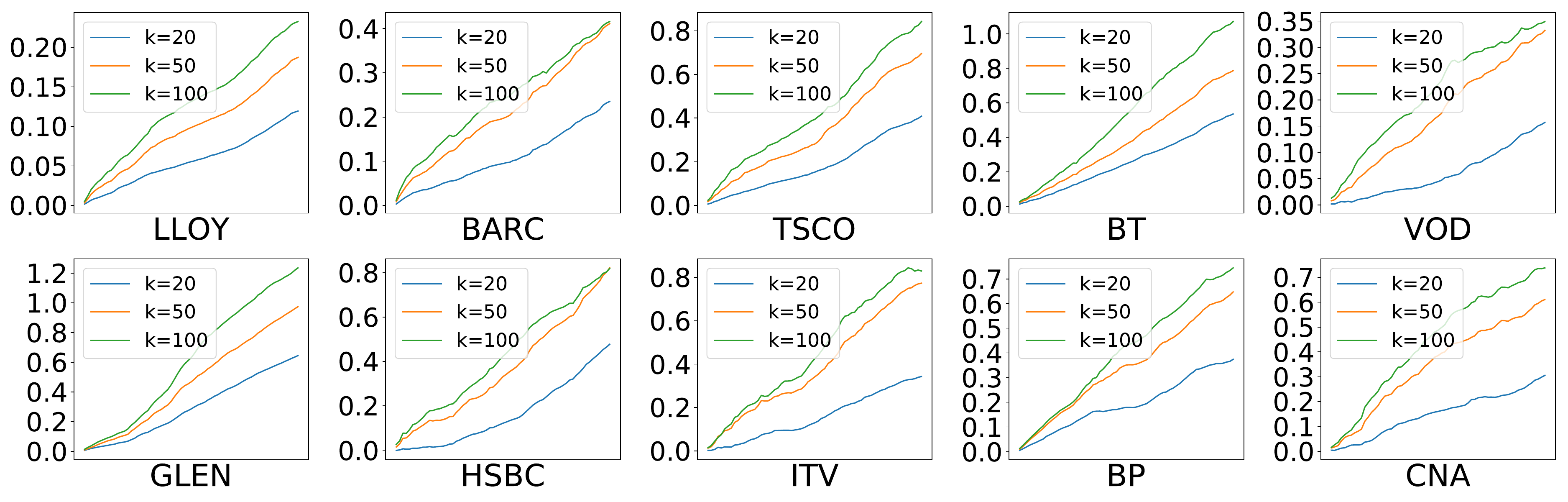}
\caption{ Normalised cumulative profits for test periods for different stocks and prediction horizons $(k)$. Profits are in GBX($=\text{GBP}/100$).}
\label{fig:cum_profit}
\end{figure*}

Figure~\ref{fig:boxplot_profit} presents the boxplots for normalised daily profits (profits divided by number of trades in that day) for different stocks and prediction horizons. We use a $t$-test to check if the profits are statistically greater than 0. The $t$-statistics is essentially the same as Sharpe ratios but a more consistent evaluation metric for high frequency trading. Figure~\ref{fig:cum_profit} shows the cumulative profits across the testing period. We can observe consistent profits and significant $t$-values over the testing period for all stocks. Although we obtain worse accuracy for longer prediction horizons, the cumulative profits are actually higher as a more robust signal is generated.

\subsection{Sensitivity Analysis}
Trust and risk are fundamental in any financial application. If we take actions based on predictions, it is always important to understand the reasons behind those predictions. Neural networks are often considered as ``black boxes'' which lack  interpretability. However, if we understand the relationship between the inputs' components (e.g. words in text, patches in an image) and the model's prediction, we can compare those relationships with our domain knowledge to decide if we can accept or reject a prediction. 

The work of \cite{ribeiro2016should} proposes a method, which they call LIME, to obtain such explanations. In our case, we use LIME to reveal components of LOBs that are most important for predictions and to understand why the proposed model DeepLOB works better than other network architectures such as CNN-I \cite{tsantekidis2017forecasting}. LIME uses an interpretable model to approximate the prediction of a complex model on a given input. It locally perturbs the input and observes variations in the model's predictions, thus providing some measure of information regarding input importance and sensitivity.

\begin{figure}[!t]
\centering
\includegraphics[width=3.25in, height=3in]{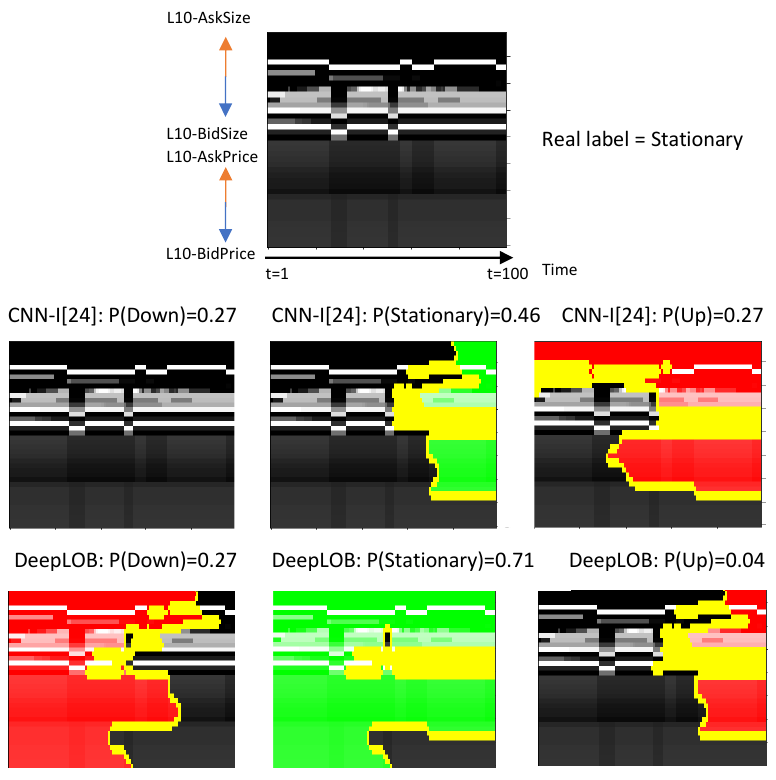}
\caption{LIME plots. x-axis represents time stamps and y-axis represents levels of the LOB, as labelled in the top image. \textbf{Top:} Original image. \textbf{Middle:} Importance regions for CNN-I \cite{tsantekidis2017forecasting}. \textbf{Bottom:} Importance regions for DeepLOB model. Regions supportive for prediction are shown in green, and regions against in red. The boundary is shown in yellow.}
\label{fig:limeplot}
\end{figure}

Figure~\ref{fig:limeplot} presents an example that shows how DeepLOB and CNN-I \cite{tsantekidis2017forecasting} react to a given input. In the figure we show the top 10 areas of pros (in green) and cons (in red) for the predicted class (yellow being the boundary). Not coloured areas represent the components of inputs that are less influential on the predicted results or ``unimportant''.  We note that  most components of the input are inactive for CNN-I \cite{tsantekidis2017forecasting}. We believe that this is due to two max-pooling layers used in that architecture. Because \cite{tsantekidis2017forecasting} used large-size filters in the first convolutional layer, any representation deep in the network actually represents information gleaned from a large portion of inputs. Our experiments applying LIME to many examples indicate this observation is a common feature. 

\section{Conclusion}\label{conclusion}

In this paper, we introduce the first hybrid deep neural network to predict stock price movements using high frequency limit order data. Unlike traditional hand-crafted models, where features are carefully designed, we utilise a CNN and an Inception Module to automate feature extraction and use LSTM units to capture time dependencies. 

The proposed method is evaluated against several baseline methods on the FI-2010 benchmark dataset and the results show that our model performs better than other techniques in predicting short term price movements. We further test the robustness of our model by using one year of limit order data from the LSE with a testing period of three months. An interesting observation from our work is that the proposed model generalises well to instruments that did not form part of the training data. This suggests the existence of universal features that are informative for price formation and our model appears to capture these features, learning from a large data set including several instruments. A simple trading simulation is used to further test our model and we obtain good profits that are statistically significant.

To go beyond the often-criticised ``black box'' nature of deep learning models, we use LIME, a method for sensitivity analysis, to indicate the components of inputs that contribute to predictions. A good understanding of the relationship between the input's components and the model's prediction can help us decide if we can accept a prediction. In particular, we see how the information of prices and sizes on different levels and horizons contribute to the prediction which is in accordance with our econometric understanding. 

In a recent extension of this work we have modified the DeepLOB model to use Bayesian neural networks \cite{zhang2018bdlob}. This allows to provide uncertainty measures on the network's outputs which for example can be used to upsize positions as demonstrated in \cite{zhang2018bdlob}. 

In subsequent continuations of this work we would like to investigate more detailed trading strategies, using Reinforcement Learning, which are based on the feature extraction performed by DeepLOB.

\section*{Acknowledgements}
The authors would like to thank members of Machine Learning Research Group at the University of Oxford for their helpful comments on drafts of this paper. We are most grateful to the Oxford-Man Institute of Quantitative Finance, who provided limit order data and other support. Computation for our work was supported by Arcus Phase B and JADE HPC at the University of Oxford and Hartree national computing facilities, U.K. We also thank the Royal Academy of Engineering U.K. for their support.

\ifCLASSOPTIONcaptionsoff
  \newpage
\fi



\end{document}